\documentclass[runningheads]{llncs}
\usepackage[T1]{fontenc}
\usepackage{graphicx}
\usepackage{booktabs}
\usepackage{amsmath}
\usepackage{amssymb}
\usepackage{textcomp}
\usepackage{booktabs,tabularx}
\usepackage{booktabs}
\usepackage[misc]{ifsym}
\newcommand{\corr}{(\Letter)}
\usepackage{mwe}
\usepackage{tabularx}
\usepackage{siunitx}
\usepackage{ragged2e}
\usepackage{array}
\usepackage{pgfplots}
\pgfplotsset{compat=1.18}
\usepackage{enumitem}
\usepackage{threeparttable}

\tocauthor{Weisheng Li, Chuqiao Huang,
  Pengcheng Li, Zhengchao Peng, Qiang Xiao,
  Zhongqian Xie, Qiang Huang, Chuanjiang Luo}
\toctitle{PIANO: Personalized Reranking via Information Aggregation Node for Music Search Optimization}

\begin{document}
\title{PIANO: Personalized Reranking via Information Aggregation Node for Music Search Optimization}
\titlerunning{PIANO: Personalized Re-ranking for Music Search}

\author{
  Weisheng Li\inst{1}\textsuperscript{*} \and
  Chuqiao Huang\inst{1}\textsuperscript{*} \and
  Pengcheng Li\inst{1} \and
  Zhengchao Peng\inst{1} \and
  Qiang Xiao\inst{1} \corr \and
  Zhongqian Xie\inst{1} \and
  Qiang Huang\inst{1} \and
  Chuanjiang Luo\inst{1}
}

\authorrunning{W. Li et al.}

\institute{
  NetEase Cloud Music, Hangzhou, China \\
  \email{\{liweisheng01, huangchuqiao01, lipengcheng06, pengzhengchao,
        hzxiaoqiang, hzxiezhongqian, huangqiang05, luochuanjiang03\}@corp.netease.com}
}

\maketitle

\begingroup
\renewcommand{\thefootnote}{\fnsymbol{footnote}}
\footnotetext{* Equal contribution.}
\endgroup
\begin{abstract}
Unlike short-video content, music tracks have long lifecycles and lasting value. Effective music search re-ranking must therefore align the user's current query with long-term preferences while jointly optimizing Click-Through Rate (CTR) and Conversion Rate (CVR). However, existing methods suffer from two limitations: (1) sequential methods rely on item-interaction history and therefore cannot use historical search queries to tell which past preferences match the user's current search intent; (2) most listwise models optimize a single objective (e.g., CTR only), and conventional multi-objective methods balance click and conversion at the item level, ignoring how these trade-offs play out across the whole ranked list. To address these limitations, we propose PIANO, a personalized listwise re-ranking framework with two key components: (i) the Query-Driven Interest Refiner (QDIR) uses cross-attention over historical queries to align past intents with the current one; (ii) the Information Aggregation Node (IAN), a learnable [CLS]-style token, aggregates the candidate list and predicts CTR/CVR at the list level. Extensive experiments on public and industrial datasets show consistent gains over strong baselines. In online A/B tests on NetEase Cloud Music, a leading music streaming platform, PIANO achieves statistically significant improvements in CTR (+0.62\%) and CVR (+4.45\%).


\end{abstract}

\section{Introduction}
Music streaming platforms depend heavily on search to link users with specific content. Unlike short-video feeds that prioritize short-term freshness, music tracks typically have long lifecycles and lasting artistic value. This presents a unique challenge for search re-ranking: systems must align users’ immediate query intent with long-term listening preferences to ensure sustained satisfaction. Moreover, industrial scenarios require balancing user experience and commercial goals, calling for joint optimization of Click-Through Rate (CTR) and Conversion Rate (CVR).

Despite the critical role of search, current re-ranking frameworks face significant hurdles in this domain.
Traditional listwise approaches, such as DLCM~\cite{ai2018dlcm}, PRM~\cite{pei2019prm}, and SetRank~\cite{pang2020setrank}, excel at modeling item interactions within the candidate list but either do not explicitly model user representations (DLCM, SetRank) or rely on a pre-computed user embedding (PRM), leaving historical query semantics underutilized.
Meanwhile, two-stage frameworks like PRS~\cite{feng2021prs} improve efficiency but are not designed to capture the semantic evolution of user intent across search queries.
This leads to two fundamental gaps in music search scenarios:

First, intent alignment is insufficient.
Most sequential models (e.g., DIN~\cite{zhou2018din}, SASRec~\cite{kang2018sasrec}) encode behavior sequences but ignore the rich semantic signals in historical queries.
While recent query-aware works~\cite{he2022query} exist, they typically rely on static query representations.
For example, a user frequently listening to \textit{fast-paced pop} might suddenly search for \textit{calm piano for studying}.
Without explicitly modeling the query sequence, the system cannot filter out contextually inconsistent historical preferences, leading to irrelevant recommendations.

Second, effective list-level multi-objective optimization remains challenging. While recent work~\cite{cao2025sortgen} explores this direction via generative modeling, discriminative frameworks that explicitly use historical search queries to align user intent are still scarce.
Existing listwise models are predominantly designed for single-objective tasks (e.g., CTR only) or rely on item-level proxies for multi-objective goals.
Conventional multi-objective methods often optimize conflicts (e.g., CTR vs. CVR) at the item level, ignoring how these trade-offs manifest across the entire ranked list.
For instance, an item-level multi-objective ranker may push a paid-content song to the top of the list because it has a high CVR score, even though placing it among a list of free indie tracks may hurt the overall click-through rate of the list. Such cross-item trade-offs cannot be captured by item-wise scoring alone.
Directly applying these methods to music search thus fails to capture the holistic utility of the slate, resulting in suboptimal balancing of business goals.

To bridge these gaps, we propose \textbf{PIANO}, i.e., \textbf{P}ersonalized reranking via \textbf{I}nformation \textbf{A}ggregation \textbf{N}ode for music search \textbf{O}ptimization---a listwise re-ranking framework with two key innovations:
(1) the Query-Driven Interest Refiner (QDIR) that leverages historical query sequences to dynamically refine long-term preferences via cross-attention;
and (2) the Information Aggregation Node (IAN), a learnable  \texttt{[CLS]}-style token initialized by QDIR to capture global list context for direct multi-objective supervision.

Our main contributions are summarized as follows:
\begin{itemize}[leftmargin=1.2em,itemsep=0.2ex,topsep=0.2ex]
    \item Query-Aware Interest Modeling: We propose QDIR to dynamically align long-term preferences with short-term query intent using historical query sequences.
    \item List-Level Multi-Objective Optimization: We design IAN, a \texttt{[CLS]}-style token that aggregates list context for direct multi-objective prediction, avoiding item-level proxy optimization.
    \item Industrial Validation: We validate PIANO through extensive offline experiments on public datasets and online A/B tests on a leading music streaming platform, yielding significant gains in CTR (+0.62\%) and CVR (+4.45\%).
\end{itemize}

\section{Related Work}

\paragraph{Sequential User Interest Modeling.}
Sequential recommenders capture dynamic user intent from interaction histories.
Self-attention models like SASRec~\cite{kang2018sasrec} model long-range dependencies, while industrial systems like BST~\cite{chen2019bst}, DIN~\cite{zhou2018din}, and SIM~\cite{pi2020sim} integrate behavior sequences with rich features.
Recent works like MIR~\cite{xi2022mir} and RAISE~\cite{lin2022raise} further enrich personalized reranking with multi-level set-to-list interactions and review-based intention modeling, respectively.
He et al.~\cite{he2022query} incorporate queries into sequential modeling via heterogeneous sequences, but rely on static query representations (e.g., average pooling) without cross-attention against historical queries, which may not fully capture the rapid intent shifts specific to music search re-ranking.
Our QDIR employs query-aware cross-attention to dynamically refine long-term preferences based on the current query.

\paragraph{Listwise Re-ranking and Contextual Modeling.}
Multi-stage recommenders increasingly adopt listwise models to capture cross-item dependencies, crucial for tasks like music search coherence~\cite{gong2020contextualPersonalizedRR}.
Existing approaches generally fall into one-stage scoring and two-stage permutation frameworks.
\textbf{One-stage models} refine candidate scores by encoding the list, evolving from RNNs (DLCM~\cite{ai2018dlcm}, Seq2Slate~\cite{bello2018seq2slate}) to self-attention mechanisms (PRM~\cite{pei2019prm}, SetRank~\cite{pang2020setrank}, RankFormer~\cite{buyl2023rankformer}).
\textbf{Two-stage frameworks} decouple generation and evaluation to balance efficiency and effectiveness (PRS~\cite{feng2021prs}, PIER~\cite{shi2023pier}, CLIG~\cite{yang2025comprehensive}).
To capture global context, these methods typically rely on \texttt{[CLS]}-style tokens or set-to-list attention, yet such nodes are rarely supervised by true list-level outcomes.
While RankFormer~\cite{buyl2023rankformer} also augments a listwise Transformer with a \texttt{[CLS]} token supervised by a listwide objective, its listwide label is synthesized as the maximum item label rather than a true per-list outcome, and the token itself is a generic learnable parameter; in contrast, IAN is initialized by QDIR's query-conditioned user interest and is directly supervised by real per-list CTR/CVR rates under a multi-objective formulation.
Furthermore, regarding user context integration, while works like PEAR~\cite{li2022pear} incorporate history, they encode user history with item-history alone and lack an explicit query-conditioned alignment of past preferences to the current intent.
This is suboptimal for dynamic scenarios like music search.
In contrast, our approach introduces an explicitly supervised global node (IAN) and employs query-aware cross-attention to dynamically align intent with preferences.

\paragraph{Multi-objective Re-ranking.}
Multi-objective re-ranking balances conflicting goals such as CTR and CVR.
Item-level multi-task models like ESMM~\cite{ma2018entire}, MMoE~\cite{ma2018modeling}, and PLE~\cite{tang2020progressive} jointly predict multiple objectives per item but ignore cross-item dependencies at re-ranking time.
Beyond item-level approaches, Nguyen et al.~\cite{nguyen2017multi} use Kendall tau regularization for multi-stakeholder ranking, and CMR~\cite{chen2023cmr} adjusts task weights via hypernetworks.
PRECTR~\cite{chen2025prectr} couples relevance matching with CTR prediction via conditional probability fusion and a batch-wise listwise consistency regularization, yet its supervision is still defined over per-item click labels rather than per-list outcomes.
SORT-Gen~\cite{cao2025sortgen} pursues list-level multi-objective optimization via autoregressive generation with ordered-regression supervision; in contrast, PIANO adopts a discriminative framework with an explicitly supervised \texttt{[CLS]}-style aggregator (IAN) that predicts per-list CTR/CVR rates in a single forward pass, enabling efficient end-to-end joint optimization.

\section{Methodology}
\subsection{Overall Architecture}
\begin{figure*}[t]
    \centering
    \includegraphics[width=1.0\textwidth]{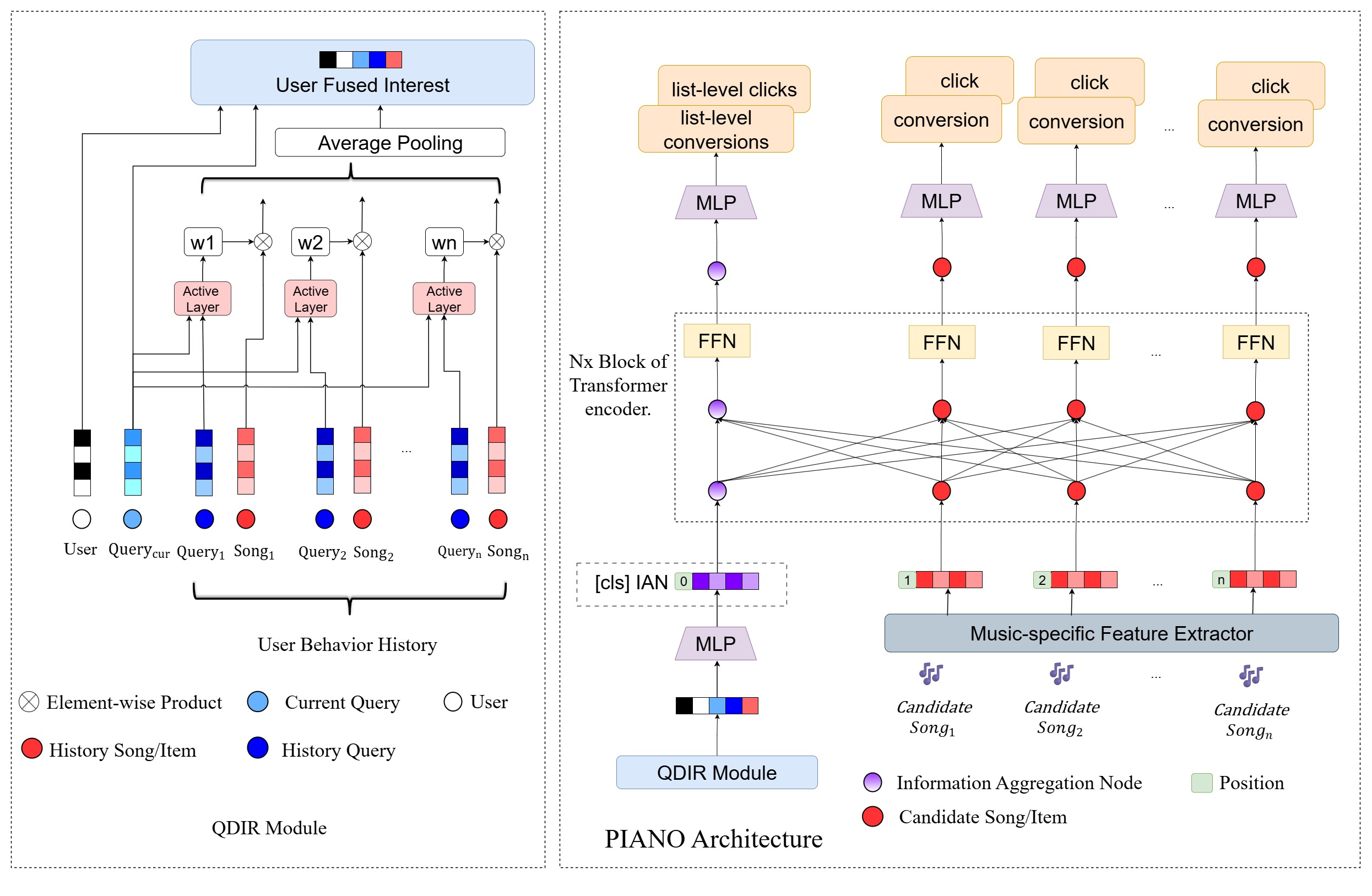}
    \caption{Overall architecture of PIANO.
    The QDIR module extracts user interest representations from query-item interaction history.
    The List-level Information Aggregation Node (IAN) with fixed position encoding is prepended to aggregate global information,
    which then interacts with candidate items via cross-attention for list-level prediction.}
    \label{fig:piano_architecture}
\end{figure*}

As illustrated in Figure~\ref{fig:piano_architecture}, PIANO comprises two core components: the Query-Driven Interest Refiner (QDIR) and the Information Aggregation Node (IAN).

Given input signals (query, user history, and candidate items), PIANO first extracts query-conditioned user interests via QDIR, which then initialize the IAN token prepended to the candidate list. Through a Transformer encoder, the model jointly optimizes both item-level and list-level business metrics. In the following subsections, we detail each component and their interactions.

\subsection{Problem Definition}
A multi-stage search/recommendation pipeline typically comprises matching, ranking, and re-ranking. Denote the result from ranking stages as \(I=(i_1,\ldots,i_M)\).
Let \(\Pi_M\) be the set of all permutations (all possible orderings) of \(\{1,\ldots,M\}\).
For \(\pi\in\Pi_M\), write \(\pi(I)=(i_{\pi(1)},\ldots,i_{\pi(M)})\).
Then the re-ranker selects
\begin{equation}
\label{eq:mo-obj}
    \pi^{*}\in \operatorname*{arg\,max}_{\pi \in \Pi_M}
\;\lambda\, U_{\mathrm{clk}}(\pi(I),C) + (1-\lambda)\, U_{\mathrm{conv}}(\pi(I),C)
\end{equation}

where \(\lambda\in[0,1]\) controls the trade-off between the click- and conversion-oriented list-level objectives. In our music search scenario, a \emph{click} denotes a user tapping a song in the returned result list, while a \emph{conversion} refers to a paid membership subscription triggered after such interaction---a key revenue indicator for music streaming platforms. Specifically, \(U_{\mathrm{clk}}(\cdot)\) and \(U_{\mathrm{conv}}(\cdot)\) denote the \emph{list-wise utility functions}.
\(U_{\mathrm{clk}}\) typically represents the \emph{expected average clicks} over the list,
while \(U_{\mathrm{conv}}\) represents the \emph{expected average conversions}.
Both are computed by aggregating item-level predictions across the entire ranked list \(\pi(I)\),
thereby necessitating a model capable of capturing global list dependencies.

We adopt a weighted-sum scalarization; by varying \(\lambda\),
the system sweeps operating points that approximate the Pareto frontier in
practice~\cite{jannach2023_mors}.

\subsection{Music-specific Representation Model}
\label{sec:music_feature_extractor}
\begin{figure}[t]
    \centering
    \includegraphics[width=1.0\textwidth]{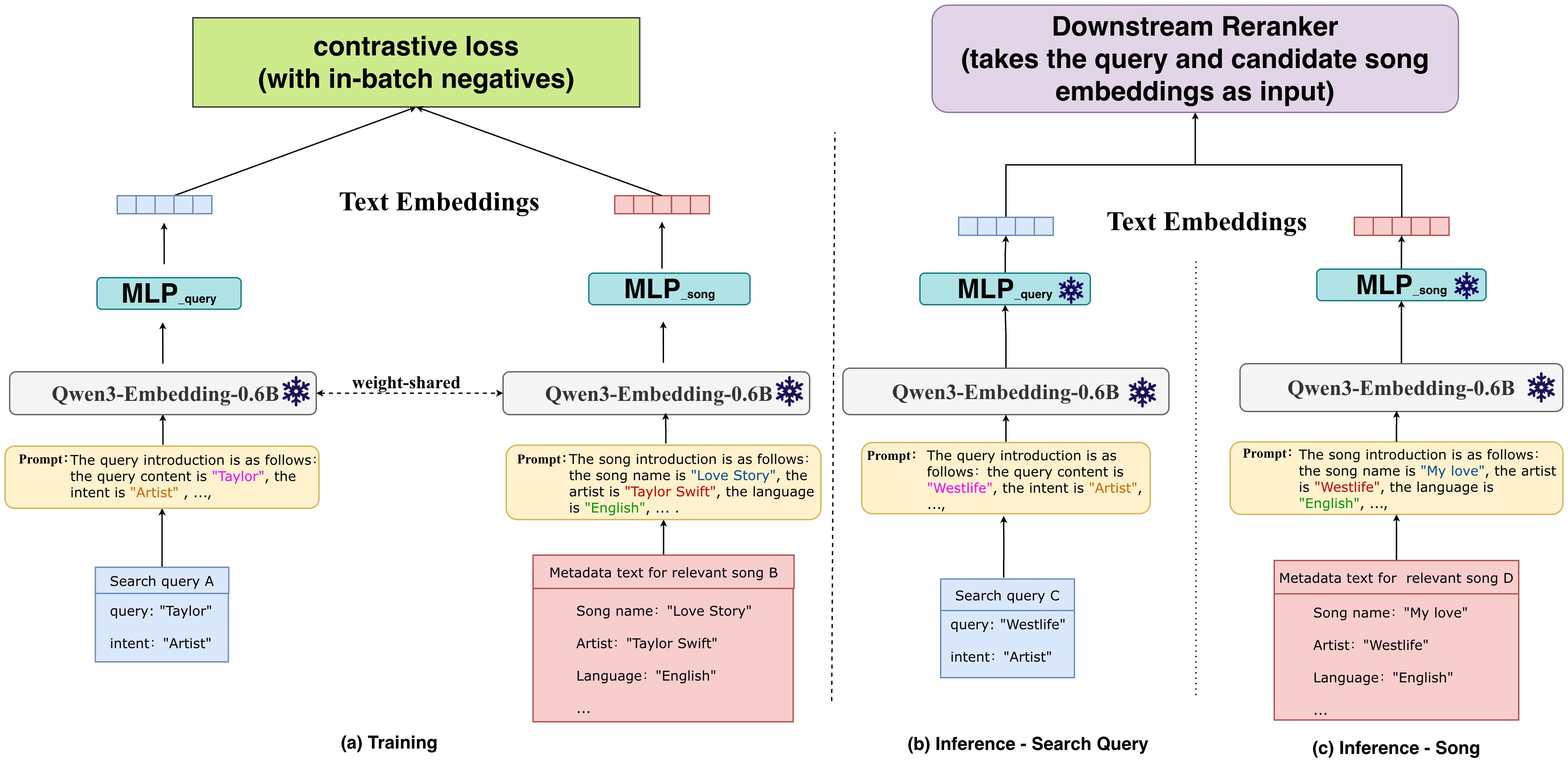}
    \caption{Pretrained music-specific representation model.
    (a) \textbf{Training}: a frozen Qwen3-Embedding-0.6B encoder is weight-shared between the query and song towers; two tower-specific MLP projection heads ($\mathrm{MLP}_{query}$, $\mathrm{MLP}_{song}$) produce text embeddings, optimized with a contrastive loss using in-batch negatives.
    (b)(c) \textbf{Inference}: both the encoder and the projection heads are frozen.
    Song embeddings are pre-computed offline, while the query embedding is computed online per search request.
    Both feed into the downstream re-ranker.}
    \label{fig:unified-embedding}
\end{figure}

To capture music-specific semantics, we employ a representation model based on Qwen3-Embedding-0.6B~\cite{zhang2025qwen3}, fine-tuned on in-domain query--song pairs with task-specific prompts.
The dual-tower design and the training/inference pipeline are illustrated in Figure~\ref{fig:unified-embedding}.
Compared to general-purpose encoders, this yields embeddings better tailored to music search and decouples encoder cost from online latency, since song embeddings are pre-computed offline.

\subsection{Query-Driven Interest Refiner}
\label{sec:qdir}
As shown in Figure~\ref{fig:piano_architecture}, the \textbf{Query-Driven Interest Refiner (QDIR)} aims to align a user’s long-term preferences with the short-term intent expressed by the current query. It consists of two components: \textit{Current Intent Representation} and \textit{Historical Query-aware Preference Refinement}.
\paragraph{Current Intent Representation.}
We first denote the raw current query as $q_{\mathrm{cur}}$.
To obtain its semantic representation, we embed the current query using the method demonstrated in Section~\ref{sec:music_feature_extractor}:
\begin{equation}
q_{\mathrm{cur}}^{\mathrm{emb}} \;=\; \phi_{\mathrm{music}}(q_{\mathrm{cur}})
\end{equation}
We then form the short-term interest by a learnable linear projection:
\begin{equation}
r_{\mathrm{short}}^{\mathrm{emb}} \;=\; W^{E}\, q_{\mathrm{cur}}^{\mathrm{emb}}
\end{equation}
where \(W^{E}\) is a learnable projection initialized to the identity, so that training starts from \(r_{\mathrm{short}}^{\mathrm{emb}} = q_{\mathrm{cur}}^{\mathrm{emb}}\) and gradually adapts the current-query embedding into the fusion space used by the cross-attention below.

\paragraph{Historical Query-aware Preference Refinement.}
To filter past preferences by the current query, we apply cross-attention~\cite{vaswani2017attention} with $q_{\mathrm{cur}}^{\mathrm{emb}}$ as the query, the historical query sequence $q_{\mathrm{seq}}^{\mathrm{emb}}=[q_{1}^{\mathrm{emb}}, \ldots, q_{L}^{\mathrm{emb}}]$ as keys, and the items interacted under each historical query $s_{\mathrm{seq}}^{\mathrm{emb}}=[s_{1}^{\mathrm{emb}}, \ldots, s_{L}^{\mathrm{emb}}]$ as values:
\begin{equation}
r_{\mathrm{long}}^{\mathrm{emb}} = \operatorname{softmax}\!\left(\frac{Q K^{\top}}{\sqrt{d}}\right) V,
\end{equation}
where $Q = W^{Q} q_{\mathrm{cur}}^{\mathrm{emb}}$, $K = W^{K} q_{\mathrm{seq}}^{\mathrm{emb}}$, $V = W^{V} s_{\mathrm{seq}}^{\mathrm{emb}}$, and $d$ is the hidden dimension. The attention weights re-weight the historical clicked items by their relevance to the current query.

\paragraph{Fusion of Long- and Short-term Interests.}
We integrate the refined long-term preference $r_{\mathrm{long}}^{\mathrm{emb}}$, the short-term intent embedding $r_{\mathrm{short}}^{\mathrm{emb}}$, and the user profile to form a unified, query-conditioned user interest representation:
\begin{equation}
r_{\mathrm{user}}^{\mathrm{emb}}
= \mathrm{LayerNormalization}\Big( W_f \cdot \left[\operatorname{concat}(r_{\mathrm{long}}^{\mathrm{emb}},\, r_{\mathrm{short}}^{\mathrm{emb}}, r_{\mathrm{user\_profile}})\right] + b \Big),
\end{equation}
where $W_f$ is a projection matrix mapping to dimension \(d_{\mathrm{model}}\), \(b\) is a bias, and \(r_{\mathrm{user\_profile}}\) denotes the static user profile embedding.
This final representation \(r_{\mathrm{user}}^{\mathrm{emb}}\) serves as the initialization signal for the IAN token in the subsequent re-ranking stage.

\subsection{Information Aggregation Node (IAN)}
\label{sec:ian}
\paragraph{Motivation and Design.}
Inspired by \texttt{[CLS]} tokens in BERT~\cite{devlin2019bert} and ViT~\cite{dosovitskiy2021vit}, we prepend a learnable \texttt{[CLS]}-style token---the \textbf{Information Aggregation Node (IAN)}---to the candidate list (Figure~\ref{fig:piano_architecture}, right). Unlike static pooling (e.g., mean) that ignores item interactions, IAN leverages self-attention to capture list-wise dependencies and to learn a list-level utility that goes beyond a simple sum of item scores.

\paragraph{Initialization and Positional Encoding.}
The IAN token is initialized using the query-conditioned interest vector from QDIR.
Let \(g = r_{\mathrm{user}}^{\mathrm{emb}} \in \mathbb{R}^{d_{\mathrm{model}}}\) denote this vector.
As shown in Figure~\ref{fig:piano_architecture}, raw candidate songs are first processed by the music-specific representation model (Section~\ref{sec:music_feature_extractor}) to obtain item representations \(h_{1:M} \in \mathbb{R}^{d_{\mathrm{model}}}\).
We project \(g\) through an MLP to obtain the raw IAN representation:
\begin{equation}
\label{eq:ian-init}
h_{0}^{\mathrm{raw}} \;=\; \mathrm{MLP}_{\mathrm{init}}(g).
\end{equation}
We then add positional encodings to all positions:
\begin{equation}
\label{eq:pe-add}
\tilde{h}_{j} \;=\;
\begin{cases}
h_{0}^{\mathrm{raw}} + \mathrm{PE}(0), & \text{if } j = 0, \\
h_{j} + \mathrm{PE}(j), & \text{if } j \in \{1, \ldots, M\},
\end{cases}
\end{equation}
where \(\mathrm{PE}(0)\) anchors the IAN to a fixed reference position, and \(\mathrm{PE}(j)\) for \(j \ge 1\) retains order sensitivity for item scoring.

\subsection{Contextualized Transformer Encoder}
\label{sec:list-encoder}
\paragraph{Architecture.}
The positionally-encoded sequence \([\tilde{h}_{0}, \tilde{h}_{1}, \ldots, \tilde{h}_{M}]\) is fed into a stack of \(N_x = 2\) pre-norm Transformer encoder blocks~\cite{vaswani2017attention}:
\begin{equation}
\label{eq:ian-ctx}
\big[ h_{0}',\, h_{1}',\, \ldots,\, h_{M}' \big]
\;=\; \mathrm{Transformer}\big(\big[ \tilde{h}_{0},\, \tilde{h}_{1},\, \ldots,\, \tilde{h}_{M} \big]\big).
\end{equation}
Each block consists of multi-head self-attention (MHSA) and a position-wise feed-forward network (FFN), both equipped with residual connections and layer normalization.

\paragraph{Prediction Heads.}
After encoding, the output sequence \([\,h_{0}',h_{1}',\ldots,h_{M}'\,]\) is utilized by two jointly trained heads for \textbf{multi-task prediction} (e.g., click and conversion, as shown in the orange boxes of Figure~\ref{fig:piano_architecture}):
\begin{subequations}
\label{eq:heads}
\begin{align}
\hat{\mathbf{p}}^{(t)}_{\mathrm{list}} &= \operatorname{Sigmoid}\big(\mathrm{MLP}(h_0')\big), \label{eq:list} \\
\mathbf{p}^{(t)}_j &= \operatorname{Sigmoid}\big(\mathrm{MLP}(h'_j)\big), \label{eq:item}
\end{align}
\end{subequations}
where \(h_{0}'\) acts as a \emph{global list representation} for the list-level classifier (Eq.~\ref{eq:list}), and the item heads (Eq.~\ref{eq:item}) operate on item representations \(h'_j\) for \(j \in \{1, \ldots, M\}\).
Crucially, the global semantics aggregated by \(h_0'\) are propagated to each item representation \(h'_j\) through self-attention, enhancing their contextual awareness.

\subsection{Model Prediction and Optimization}
\label{sec:loss_function}
We train PIANO by employing two complementary loss functions to jointly optimize list-level and item-level objectives.
Note that PIANO uses three weights at different levels: \(\lambda\) (Eq.~\ref{eq:mo-obj}) for click vs.\ conversion utility; \(\alpha\) (Eq.~\ref{eq:joint}) for list- vs.\ item-level loss; \(\beta\) for click vs.\ conversion BCE within the item loss.
\paragraph{List--level loss.}
Aligned with Eq.~\eqref{eq:mo-obj}, we predict the list-level click rate and conversion rate via two BCE losses on soft labels:
\begin{equation}
\label{eq:list-loss}
L_{\mathrm{list}}
= -\frac{1}{N}\sum_{t=1}^{N} \sum_{m \in \{\text{clk}, \text{conv}\}} \left[ y_{m}^{(t)} \log \hat{p}_{m}^{(t)} + (1 - y_{m}^{(t)}) \log (1 - \hat{p}_{m}^{(t)}) \right]
\end{equation}
where \(\mathbf{y}^{(t)} = (y_{\text{clk}}^{(t)}, y_{\text{conv}}^{(t)}) \in [0,1]^2\) are the per-list rate targets, \(\hat{\mathbf{p}}^{(t)} = (\hat{p}_{\text{clk}}^{(t)}, \hat{p}_{\text{conv}}^{(t)})\) are the predicted scores, $N$ is the batch size, and the targets are defined as:
\begin{align}
\label{eq:rate-def}
y_{\text{clk}}^{(t)} &= \frac{\text{\#clicked items in list } t}{\text{length of list } t}, \\
y_{\text{conv}}^{(t)} &= \frac{\text{\#converted items in list } t}{\text{length of list } t}.
\end{align}

\paragraph{Item--level loss.}
For item labels \(y_{j}^{(\mathrm{clk})}, y_{j}^{(\mathrm{conv})}\in\{0,1\}\), we use binary cross--entropy (BCE) with a conversion weight \(\beta\ge 0\):
\begin{align}
L_{\mathrm{item}}
&= \frac{1}{N}\sum_{t=1}^{N} \frac{1}{M}\sum_{j=1}^{M}
\Big(
\ell_{\mathrm{BCE}}\!\big(y_{j}^{(\mathrm{clk})},\,p_{j}^{(\mathrm{clk})}\big)
\;+\;
\beta\,
\ell_{\mathrm{BCE}}\!\big(y_{j}^{(\mathrm{conv})},\,p_{j}^{(\mathrm{conv})}\big)
\Big)
\end{align}
\begin{align}
\ell_{\mathrm{BCE}}(y,p) &= -\,y\log p - (1-y)\log(1-p),
\end{align}
where \(\beta\) balances the conversion signal relative to click signal. Typically, we set \(\beta = 1\).

\paragraph{Joint objective.}
The final training objective is a weighted sum:
\begin{equation}
\label{eq:joint}
L \;=\; \alpha\, L_{\mathrm{list}} \;+\; (1-\alpha)\,L_{\mathrm{item}},
\end{equation}
where \(\alpha > 0\) balances list--level supervision and item-level supervision (we study sensitivity to \(\alpha\) in Section~\ref{sec:alpha_sensitivity}).
Through self-attention, the IAN aggregates global list context into \(h'_0\), which is supervised by \(L_{\mathrm{list}}\) for holistic utility prediction.
Meanwhile, item representations \(h'_j\) benefit from this global context via attention, enabling joint optimization aligned with the multi-objective target in Equation~\eqref{eq:mo-obj}.

\section{Offline Experiments}
We evaluate \textsc{PIANO} on the public Yahoo Letor benchmark~\cite{chapelle2011learning} and our Industrial Music Search Re-ranking Dataset. We first compare against strong baselines, then perform ablations and sensitivity analysis. Specifically, our aim is to address the following research questions.

\begin{itemize}[leftmargin=1.2em,itemsep=0.2ex,topsep=0.4ex]
    \item \textbf{RQ1} Does \textsc{PIANO} outperform strong re-ranking baselines across datasets and metrics?
    \item \textbf{RQ2} Do ablations confirm that \textbf{QDIR} and \textbf{IAN} each provide measurable gains?
    \item \textbf{RQ3} How does the balancing hyperparameter $\alpha$ affect overall performance?
    \item \textbf{RQ4} How does the length of the historical behavior sequence for QDIR impact ranking effectiveness?
    \item \textbf{RQ5} Does the IAN encode meaningful list-level semantics, as evidenced by diagnostics/visualization?
\end{itemize}

\subsection{Experimental Settings}
\begin{table}[t]
  \centering
  \caption{Overview of the Datasets}
  \label{tab:datasets}
  \begin{tabularx}{\linewidth}{l *{2}{>{\Centering\arraybackslash}X}}
    \toprule
    & Yahoo Letor Dataset
    & Music Search Re-ranking Dataset
    \\
    \midrule
    \#User        & -        & 4,488,871      \\
    \#Docs/Items  & 709,877  & 642,239        \\
    \#Queries     & -        & 236,191        \\
    \#Records     & 29,921   & 31,336,800     \\
    \bottomrule
  \end{tabularx}
\end{table}

\subsubsection{Datasets.} We evaluate on Yahoo! Webscope v2.0 (Yahoo Letor)~\cite{chapelle2011learning} and the Industrial Music Search Re-ranking Dataset (Table~\ref{tab:datasets}).

\textbf{Yahoo Letor Dataset.} Following~\cite{bello2018seq2slate}, we binarize relevance grades ($r \ge T_b$) and simulate position bias via $P_{\mathrm{exam}}(i) \propto 1/\mathrm{pos}(i)^{\eta}$.

\textbf{Music Search Re-ranking Dataset.} Constructed from NetEase Cloud Music, a leading music streaming platform, this dataset comprises query contexts, candidate lists, and user feedback. We filter sessions with full first-screen exposure to minimize bias. Features include metadata, behavioral statistics, and textual embeddings (Section~\ref{sec:music_feature_extractor}). The labels include item-level click and conversion signals, together with their list-wise averages as list-level CTR and CVR targets. We adopt a time-based split to prevent leakage, ensuring all methods share the same candidate pool and features.

\subsubsection{Baselines}
We compare \textsc{PIANO} with representative learning-to-rank (LTR) and listwise re-ranking baselines~\cite{Liu2009}, covering three dimensions relevant to our contributions:
(i) \emph{LTR paradigm} (pointwise/pairwise vs.\ listwise),
(ii) \emph{personalization} (with/without user-side signals), and
(iii) \emph{contextual modeling} (independent scoring vs.\ slate-aware inter-item interaction).

\textbf{SVMRank}~\cite{joachims2006svmrank} is a classic pairwise LTR
baseline. As a non-listwise, non-personalized method, it serves as
a lower bound to quantify the benefit of inter-item modeling.

\textbf{LambdaMART}~\cite{burges2010lambdamart} is a strong listwise LTR
method based on gradient-boosted decision trees, optimizing NDCG through
pairwise lambda gradients weighted by list-level metric deltas. We
include it as a representative production-grade baseline widely deployed
in industrial settings due to its low-latency inference.

\textbf{DLCM}~\cite{ai2018dlcm} is the seminal listwise re-ranker that encodes the top-$n$ slate with a GRU. As an early sequential listwise model, it serves as a contrast to the self-attention-based re-rankers introduced next.

\textbf{PRM}~\cite{pei2019prm} injects user embeddings into a
Transformer self-attention encoder to jointly model user--item and
item--item interactions. As the canonical \emph{personalized}
self-attention re-ranker, it constitutes the closest non-historical
counterpart to \textsc{PIANO}.

\textbf{SAR}~\cite{ren2023sar} is a recent listwise method that captures item--item dependencies through slate-aware contextual modeling. We include it as a contemporary listwise baseline without user-history or query-aware components.

\textbf{PEAR}~\cite{li2022pear} fuses the initial list and user-history signals via a Transformer for personalized re-ranking. As the most architecturally similar baseline to \textsc{PIANO}, PEAR relies on item-history without query signals; we use it as our key reference point for personalized listwise re-ranking.

\subsubsection{Implementation Details}
\label{sec:implementation-details}
\paragraph{Hyperparameters.} For all baselines and for \textsc{PIANO}, we keep critical hyperparameters identical across datasets to ensure fairness. We set the hidden size to \(d_{\text{model}}=128\) for both the Yahoo Letor Dataset and the Industrial Music Search Re-ranking Dataset. All models are trained with Adam~\cite{kingma2015adam}. The dropout rate is set to \(p_{\text{dropout}}=0.2\). The mini-batch size is \(32\) for the Yahoo Letor Dataset and \(256\) for the Music Search Re-ranking Dataset.

\paragraph{Evaluation Protocol.}
All experimental results are reported as mean $\pm$ standard deviation
over 10 independent runs with different random seeds to ensure statistical
reliability.

\paragraph{Feature Alignment across Datasets.}
To ensure fair comparison, we align the input representations across datasets while adapting to each dataset's available metadata.

For the Industrial Music Search Re-ranking Dataset, $s_i^{\mathrm{emb}}$ is the song embedding produced by the music-specific representation model (Section~\ref{sec:music_feature_extractor}).

For the Yahoo Letor Dataset, which lacks music metadata, we instead use a trainable ID embedding of the same width $d_{\mathrm{model}}$:
\begin{equation}
    s_i^{\mathrm{emb}} = E_{\mathrm{item}}[\mathrm{ID}(i)],
\end{equation}
where $\mathrm{ID}(\cdot)$ maps each item to its integer index in the item ID set $\mathcal{D}$, and $E_{\mathrm{item}} \in \mathbb{R}^{(|\mathcal{D}|+1) \times d_{\mathrm{model}}}$ is the embedding lookup matrix with an extra \texttt{[UNK]} row for unseen items.

In both cases, $s_i^{\mathrm{emb}}$ occupies the same $d_{\mathrm{model}}$-dimensional input slot, so all methods receive item representations of identical shape.

\paragraph{Dataset-specific adaptations.}
PEAR requires user behavior sequences unavailable in the Yahoo Letor Dataset, so we evaluate it only on the Industrial Music Search Re-ranking Dataset.
For \textsc{PIANO} on the Yahoo Letor Dataset, we disable QDIR's query-sequence component while retaining IAN, ensuring fair comparison under identical feature constraints.

\subsubsection{Evaluation Metrics}
For offline evaluation, we adopt two standard ranking metrics~\cite{jarvelin2002cumulated}: \textbf{Normalized Discounted Cumulative Gain at $k$ (NDCG@$k$)}, which weights each position by a logarithmic discount so that relevant items ranked higher contribute more to the score; and \textbf{Mean Average Precision (MAP)}, the mean over queries of the average precision computed at the positions of relevant items. Given that our industrial music search scenario displays only the top 5 results, we set $k=5$ to align with practical deployment.

\begin{table}[t]
  \centering
  \caption{Performance Comparison on Yahoo Letor Dataset}
  \label{tab:yahoo_perf}
  \begin{tabular*}{\linewidth}{@{\extracolsep{\fill}} l c c c }
    \toprule
    \textbf{Model} & \textbf{NDCG@5} & \textbf{NDCG@10} & \textbf{MAP} \\
    \midrule
    PIANO   & \textbf{0.6700$\pm$0.0008} & \textbf{0.7225$\pm$0.0006} & \textbf{0.6516$\pm$0.0005} \\
    PRM           & \underline{0.6676$\pm$0.0008} & \underline{0.7204$\pm$0.0008} & \underline{0.6494$\pm$0.0006} \\
    SAR           & 0.6539$\pm$0.0028 & 0.7107$\pm$0.0020 & 0.6386$\pm$0.0023 \\
    DLCM          & 0.6530$\pm$0.0019 & 0.7044$\pm$0.0017 & 0.6353$\pm$0.0016 \\
    LambdaMART    & 0.6388$\pm$0.0002 & 0.6992$\pm$0.0003 & 0.6153$\pm$0.0001 \\
    SVM           & 0.5943$\pm$0.0001 & 0.6556$\pm$0.0002 & 0.5768$\pm$0.0001 \\
    \bottomrule
  \end{tabular*}
\end{table}

\begin{table}[t]
  \centering
  \caption{Performance Comparison on Industrial Music Search Re-ranking Dataset}
  \label{tab:industry_perf}
  \begin{tabular*}{\linewidth}{@{\extracolsep{\fill}} l c c c }
    \toprule
    \textbf{Model} & \textbf{NDCG@3} & \textbf{NDCG@5} & \textbf{MAP} \\
    \midrule
    PIANO   & \textbf{0.8296$\pm$0.0002} & \textbf{0.8606$\pm$0.0002} & \textbf{0.8100$\pm$0.0003} \\
    PEAR          & \underline{0.8277$\pm$0.0001} & \underline{0.8586$\pm$0.0002} & \underline{0.8081$\pm$0.0002} \\
    PRM           & 0.8098$\pm$0.0006 & 0.8433$\pm$0.0006 & 0.7868$\pm$0.0008 \\
    SAR           & 0.8062$\pm$0.0002 & 0.8426$\pm$0.0001 & 0.7840$\pm$0.0002 \\
    DLCM          & 0.8042$\pm$0.0006 & 0.8398$\pm$0.0004 & 0.7822$\pm$0.0006 \\
    LambdaMART    & 0.7767$\pm$0.0016 & 0.8247$\pm$0.0017 & 0.7622$\pm$0.0012 \\
    SVM           & 0.7640$\pm$0.0014 & 0.8188$\pm$0.0012 & 0.7542$\pm$0.0016 \\
    \bottomrule
  \end{tabular*}
\end{table}

\subsection{Overall Performance (RQ1)}
Tables~\ref{tab:yahoo_perf} and \ref{tab:industry_perf} report the performance on both datasets. \textbf{PIANO} consistently outperforms all baselines across all metrics.

On the Yahoo Letor Dataset, \textsc{PIANO} achieves consistent gains over the strongest neural baseline (PEAR), validating that QDIR and IAN provide measurable benefits even with short candidate lists. The advantages are more pronounced on the Industrial Music Search Re-ranking Dataset, where \textsc{PIANO} significantly surpasses PEAR and other baselines. These results confirm that fusing query-conditioned interest modeling with explicit list-level multi-objective optimization improves re-ranking quality in real-world scenarios.

\subsection{Ablation Study (RQ2)}
\label{sec:ablation_study}
We conduct ablation studies on the Industrial Music Search Re-ranking Dataset to quantify the contribution of each component (Table~\ref{tab:ablation_study}).

\paragraph{Effect of QDIR.}
Removing \textbf{QDIR} (replaced by query-agnostic pooling) leads to consistent drops in NDCG@$k$ and MAP.
This confirms that aligning long-term behavioral signals with the current query is essential for effective personalization, as it filters noise and captures intent shifts.

\paragraph{Effect of IAN.}
Ablating \textbf{IAN} and the list-level head uniformly degrades performance across all metrics.
This validates the necessity of list-level multi-objective supervision, which provides global context complementary to item-wise signals.

\begin{table}[t]
  \centering
  \caption{Ablation Study of PIANO on Industrial Music Search Re-ranking Dataset}
  \label{tab:ablation_study}
  \begin{tabularx}{\linewidth}{@{} X *{3}{>{\centering\arraybackslash}X} @{}}
    \toprule
    \textbf{Model} & \textbf{NDCG@3} & \textbf{NDCG@5} & \textbf{MAP} \\
    \midrule
    w/ QDIR, w/ IAN    & \textbf{0.8296$\pm$0.0002} & \textbf{0.8606$\pm$0.0002} & \textbf{0.8100$\pm$0.0003} \\
    w/o QDIR, w/ IAN      & 0.8108$\pm$0.0013 & 0.8442$\pm$0.0011 & 0.7880$\pm$0.0015 \\
    w/ QDIR, w/o IAN       & 0.8198$\pm$0.0002 & 0.8507$\pm$0.0002 & 0.8002$\pm$0.0003 \\
    w/o QDIR, w/o IAN      & 0.8103$\pm$0.0014 & 0.8437$\pm$0.0013 & 0.7874$\pm$0.0018 \\
    \bottomrule
  \end{tabularx}
\end{table}

\subsection{Effect of the Balancing Parameter $\alpha$ (RQ3)}
\label{sec:alpha_sensitivity}
We investigate the impact of the balancing parameter $\alpha$ in the joint objective (Eq.~\ref{eq:joint}), which controls the trade-off between list-level and item-level supervision.

As shown in Table~\ref{tab:alpha_ablation}, performance exhibits a clear inverted-U pattern:
$\alpha\!=\!0$ (list-level only) yields suboptimal results due to insufficient item-level guidance,
while $\alpha\!=\!0.5$ achieves the best balance, indicating that global list semantics and fine-grained item signals are complementary.
Further increasing $\alpha$ to $0.75$ leads to diminishing returns, suggesting that over-emphasizing item-level loss may dilute the global aggregation capability of IAN.
This validates the necessity of joint optimization and guides our choice of $\alpha\!=\!0.5$ for all subsequent experiments.

\begin{table}[t]
  \centering
  \caption{PIANO performance with different balancing parameter $\alpha$}
  \label{tab:alpha_ablation}
  \begin{tabular*}{\linewidth}{@{\extracolsep{\fill}} l c c c }
    \toprule
    \boldmath$\alpha$\unboldmath & \textbf{NDCG@3} & \textbf{NDCG@5} & \textbf{MAP} \\
    \midrule
    0.00 & 0.8199$\pm$0.0002 & 0.8508$\pm$0.0002 & 0.8004$\pm$0.0003 \\
    0.25 & 0.8250$\pm$0.0003 & 0.8577$\pm$0.0002 & 0.8052$\pm$0.0003 \\
    0.50 & \textbf{0.8296$\pm$0.0002} & \textbf{0.8606$\pm$0.0002} & \textbf{0.8100$\pm$0.0003} \\
    0.75 & 0.8262$\pm$0.0005 & 0.8576$\pm$0.0004 & 0.8060$\pm$0.0006 \\
    \bottomrule
  \end{tabular*}
\end{table}

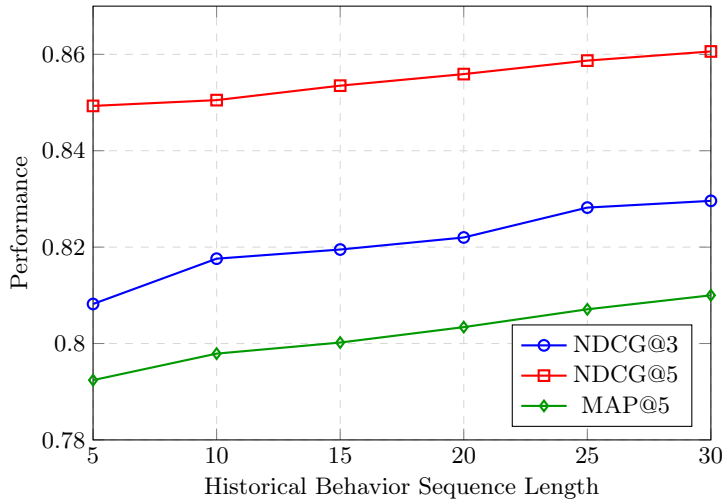
\begin{figure}[t]
\centering
\scriptsize
\begin{tikzpicture}[trim axis left, trim axis right]
\begin{axis}[
    width=0.8\linewidth,
    height=0.6\linewidth,
    xlabel={Historical Behavior Sequence Length},
    ylabel={Performance},
    xmin=5, xmax=30,
    ymin=0.78, ymax=0.87,
    xtick={5,10,15,20,25,30},
    ytick={0.78,0.80,0.82,0.84,0.86},
    legend pos=south east,
    grid=major,
    grid style={dashed, gray!30},
    tick label style={font=\small},
    label style={font=\small},
    legend style={font=\footnotesize},
]

\addplot[mark=o, blue, thick] coordinates {
(5, 0.8082)
(10, 0.8176)
(15, 0.8195)
(20, 0.8220)
(25, 0.8282)
(30, 0.8296)
};
\addlegendentry{NDCG@3}

\addplot[mark=square, red, thick] coordinates {
(5, 0.8493)
(10, 0.8505)
(15, 0.8535)
(20, 0.8559)
(25, 0.8587)
(30, 0.8606)
};
\addlegendentry{NDCG@5}

\addplot[mark=diamond, green!60!black, thick] coordinates {
(5, 0.7924)
(10, 0.7979)
(15, 0.8002)
(20, 0.8034)
(25, 0.8071)
(30, 0.8100)
};
\addlegendentry{MAP@5}
\end{axis}
\end{tikzpicture}
\caption{Impact of historical behavior sequence length on ranking performance.}
\label{fig:seq_len}
\end{figure}

\subsection{Effect of Historical Behavior List Length (RQ4)}
Figure~\ref{fig:seq_len} reports PIANO's performance under
different historical behavior list lengths. A clear monotonic pattern emerges: all metrics consistently improve as the sequence grows from 5 to 30 interactions.
These results verify that richer historical behavior sequences provide stronger long‐term preference signals for \textbf{QDIR},
while its query-driven cross-attention filters out unrelated history, so a longer sequence brings more useful signal without adding noise.

\subsection{Visualization of the IAN (RQ5)}
\label{sec:viz-ian}

\begin{figure}[t]
    \centering
    \includegraphics[width=0.9\textwidth, height=0.55\linewidth,]{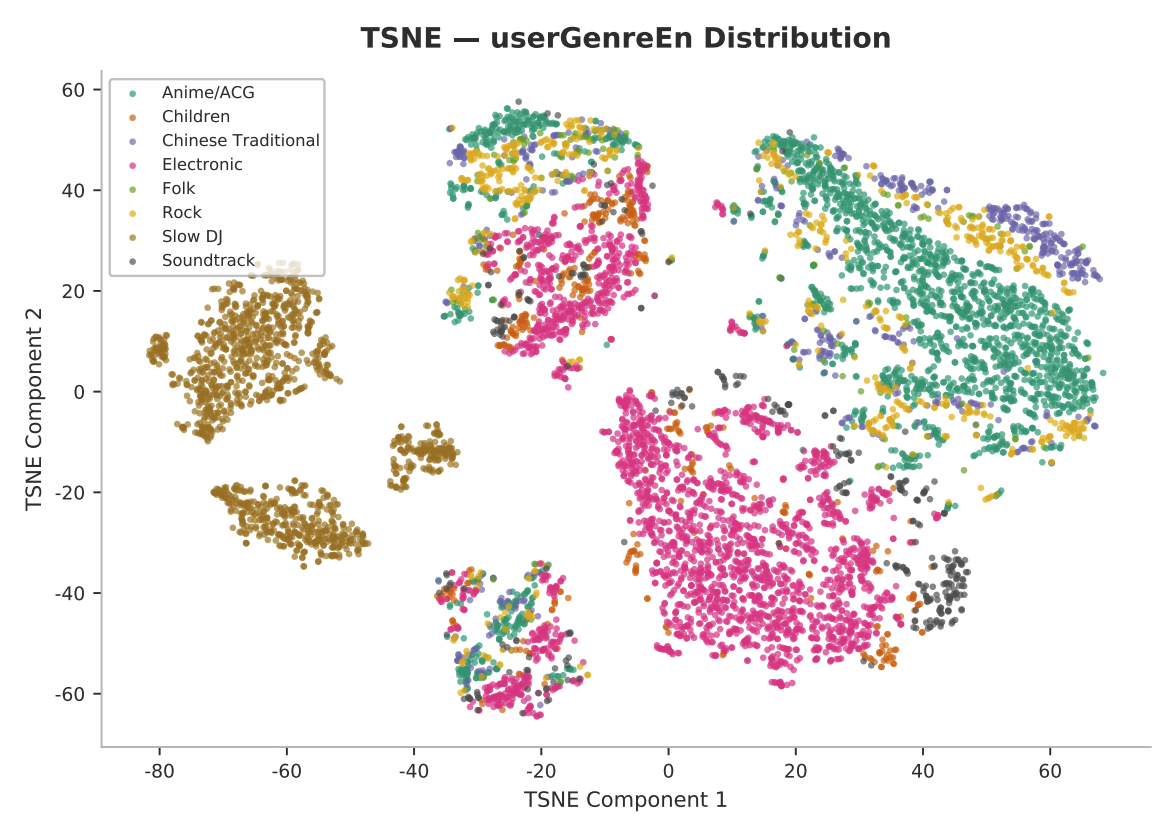}
    \caption{t-SNE of the Output Embedding of IAN after Transformer Encoder}
    \label{fig:visualization_cls_token}
\end{figure}

To assess whether the \textbf{Information Aggregation Node} (IAN) effectively encodes list-level stylistic signals, we extract the post-Transformer IAN embedding $h_0'$ for a random set of held-out users and project these vectors to two dimensions using t-SNE~\cite{maaten2008tsne}. We consider eight high-frequency genre tags---\emph{Anime/ACG}, \emph{Children}, \emph{Chinese Traditional}, \emph{Electronic}, \emph{Folk}, \emph{Rock}, \emph{Slow DJ}, and \emph{Soundtrack}---with each point in Figure~\ref{fig:visualization_cls_token} colored by the dominant genre of the candidate list.

The visualization reveals clear genre-aligned clustering: \emph{Slow DJ} forms a well-separated cluster on the left, while \emph{Electronic} and \emph{Anime/ACG} occupy distinct regions in the lower-middle and upper-right areas. Other genres exhibit discernible patterns with some overlap reflecting stylistic proximity. This demonstrates that IAN successfully aggregates global candidate list information, producing semantically meaningful representations.

Crucially, this capability directly supports our list-level multi-objective prediction task. Since user engagement varies with list composition (e.g., genre), IAN's ability to cluster stylistically different lists gives the list-level MLP head a useful basis for predicting clicks and conversions. Analogous to BERT's \texttt{[CLS]} token, IAN aggregates item-level features into a list-level summary that simultaneously feeds the list-level head and conditions item-level scoring through self-attention.

\section{Online Evaluation \& Deployment}
\subsection{Online A/B Test}
We conducted a multi-week online experiment in our industrial music search stack. Users were bucketed by stable identifiers; the treatment used \textsc{PIANO} and the
control used the pre-existing re-ranker. The trial showed statistically significant gains of \textbf{+0.62\%} CTR and
\textbf{+4.45\%} CVR (both with $p < 0.05$, two-sided). The gains were stable across weekdays/weekends and major
query categories.

\begin{table}[t]
  \centering
  \caption{Online A/B Test Results: Absolute and Relative Changes vs. Control}
  \label{tab:online_results}
  \begin{tabular*}{\linewidth}{@{\extracolsep{\fill}} l cccc }
    \toprule
    \textbf{Week} & \textbf{CTR Abs. (pp)} & \textbf{CTR Rel. (\%)} & \textbf{CVR Abs. (pp)} & \textbf{CVR Rel. (\%)} \\
    \midrule
    Week 1  & $+0.13^{*}$ \scriptsize{[0.00, 0.26]} & $+0.46^{*}$ & $+0.006^{*}$ \scriptsize{[0.003, 0.009]} & $+4.62^{*}$ \\
    Week 2  & $+0.18^{*}$ \scriptsize{[0.05, 0.31]} & $+0.66^{*}$ & $+0.007^{*}$ \scriptsize{[0.004, 0.009]} & $+4.91^{*}$ \\
    Week 3  & $+0.21^{*}$ \scriptsize{[0.05, 0.36]} & $+0.73^{*}$ & $+0.006^{*}$ \scriptsize{[0.003, 0.011]} & $+4.02^{*}$ \\
    Week 4  & $+0.17^{*}$ \scriptsize{[0.01, 0.33]} & $+0.61^{*}$ & $+0.007^{*}$ \scriptsize{[0.004, 0.013]} & $+4.23^{*}$ \\
    Avg & +0.17 & +0.62 & +0.005 & +4.45 \\
    \bottomrule
  \end{tabular*}
\begin{flushleft}
    \footnotesize
    \textit{Note:} * indicates statistical significance at $p < 0.05$ (two-tailed).
  \end{flushleft}
\end{table}

\subsection{Online Deployment}
We have deployed \textsc{PIANO} for music search at NetEase Cloud Music, a large-scale music streaming platform. The system is integrated as a post-ranking service in the existing \textit{matching$\rightarrow$ranking$\rightarrow$re-ranking} pipeline: the upstream ranker returns a list $I$, and \textsc{PIANO} produces the final order $\pi$. In steady state, 95\% of traffic is served by \textsc{PIANO} (covering millions of daily active users), while a persistent 5\% randomized holdout is routed to the baseline for long-horizon KPI tracking. We conduct periodic model refreshes using the latest user interaction logs to adapt to evolving preferences. Since launch, key business metrics have remained consistently above the control group, exhibiting sustained positive uplifts without compromising online latency requirements. This deployment demonstrates that list-aware, query-conditioned re-ranking can be operated reliably at scale in a high-traffic music search system.

\section{Conclusion}

In this paper, we identified two gaps in music search re-ranking: queries shift faster than item-history alone can capture, and list-level multi-objective signals are usually optimized via item-level proxies. PIANO addresses both with QDIR (query-driven cross-attention over historical queries) and IAN (a \texttt{[CLS]}-style token for list-level CTR/CVR prediction). Experiments on public and industrial datasets, together with a multi-week online A/B test on NetEase Cloud Music, a leading music streaming platform, show consistent gains, with $+0.62\%$ CTR and $+4.45\%$ CVR online. These findings confirm PIANO's robustness and value for deployment in real-world, latency-sensitive information retrieval systems.

\begin{credits}
\subsubsection{\discintname}
The authors have no competing interests to declare that are relevant to the content of this article.
\end{credits}

\bibliographystyle{splncs04}
\bibliography{refer}

\end{document}